# Phase-field modeling of biomineralization in mollusks and corals: Microstructure vs. formation mechanism

**László Gránásy,**[1,2,§] **László Rátkai,**[1] **Gyula I. Tóth,**[3] **Pupa U. P. A. Gilbert,**[4] **Igor Zlotnikov,**[5] **and Tamás Pusztai**[1]

[1] Institute for Solid State Physics and Optics, Wigner Research Centre for Physics, P. O. Box 49, H−1525 Budapest, Hungary
[2] Brunel Centre of Advanced Solidification Technology, Brunel University, Uxbridge, Middlesex, UB8 3PH, UK
[3] Department of Mathematical Sciences, Loughborough University, Loughborough, Leicestershire LE11 3TU, UK
[4] Departments of Physics, Chemistry, Geoscience, Materials Science, University of Wisconsin–Madison, Madison, WI 53706, USA.
[5] B CUBE–Center for Molecular Bioengineering, Technische Universität Dresden, 01307 Dresden, Germany
[§] Corresponding author: granasy.laszlo@wigner.hu

While biological crystallization processes have been studied on the microscale extensively, models addressing the mesoscale aspects of such phenomena are rare. In this work, we investigate whether the phase-field theory developed in materials science for describing complex polycrystalline structures on the mesoscale can be meaningfully adapted to model crystallization in biological systems. We demonstrate the abilities of the phase-field technique by modeling a range of microstructures observed in mollusk shells and coral skeletons, including granular, prismatic, sheet/columnar nacre, and sprinkled spherulitic structures. We also compare two possible micromechanisms of calcification: the classical route via ion-by-ion addition from a fluid state and a non-classical route, crystallization of an amorphous precursor deposited at the solidification front. We show that with appropriate choice of the model parameters microstructures similar to those found in biomineralized systems can be obtained along both routes, though the time-scale of the non-classical route appears to be more realistic. The resemblance of the simulated and natural biominerals suggests that, underneath the immense biological complexity observed in living organisms, the underlying design principles for biological structures may be understood with simple math, and simulated by phase-field theory.

**Keywords:** A1: Biomineralization, Crystallization, Calcification, Phase-field theory, Bioinspired materials

## 1. Introduction

Crystalline materials formed by solidification from the liquid state play an essential role in our civilization [1, 2]. This class of matter incorporates most of the technical alloys, polymers, minerals, drugs, food products, etc. Owing to their importance, mathematical models describing the process of crystallization under the respective conditions were and are being developed. Relying on the statistical physical description of phase transitions, the evolving numerical methods, and the ever increasing computational power, computational materials science reached the level, where knowledge based design of crystalline matter is possible for certain classes of materials (see, e.g., [2–4]). The models that address the behavior of matter during crystalline solidification range from the molecular time and length scales to the engineering scales. They include *ab initio* computations; particle-based methods like molecular dynamics, Monte Carlo, or population dynamics simulations and different types of continuum models ranging from the density functional theory of classical particles, via coarse-grained models (such as the time dependent Ginzburg-Landau, Cahn-Hilliard, and phase-field type order parameter theories that belong to the family of classical field theoretical models widely used in modeling phase transitions of various complexity), to the macroscopic continuum models applicable on engineering time- and length-scales. While this inventory allows the modelling of a substantial range of crystallization phenomena, there are complex cases, for which its use is not straightforward. Such examples are the biomorphic (inorganic) materials [5-10] that form worm-shape or arboresque morphologies by aggregation of crystalline particles, and the process of biomineralization [11-19]; i.e., the formation of hierarchically structured organic-inorganic composites in biological systems. Examples of biomineralization include the formation of mollusk shells [13, 14], skeletons of corals [15] and cell walls of diatoms [16], kidney stones [17], bones and teeth [18], and magnetite crystals in the magnetosomes of magnetotactic bacteria [19], to name a few. The materials formed by biomineralization often have surprisingly good mechanical properties owing to their hierarchical microstructure (see e.g. [13, 14]). Recent imaging and analytic methods provide detailed information on the respective microstructures, which in turn may give clues to the formation mechanism: many of these microstructures are well known from materials science (such as dendrites, spherulites, cellular, and columnar shapes, etc.) [11-15, 17-19]. This raises the possibility that with some adjustment / further development, the models developed in materials science can be used to reverse engineer the biomineralization process, and learn the pathways used by nature to create these complex structures, which may inspire new technologies for creating novel composite materials [20-25].

Recently, we explored the possibility of developing predictive mathematical models for biomorphic crystallization and for relatively simple biomineralization processes by adopting well-established methods of computational materials science and adjusting them to the circumstances as necessary [26-28]. The research done so far is confined yet to relatively simple cases of extracellular biomineralization such as mollusk shell formation [26, 27] or microstructure evolution of spherulitic structures in coral skeletons [28], but is expected to deepen the general understanding in the field, and the tools developed in the course of this research might open the way for modelling more complex cases of crystallization in biological systems such as formation of bones, kidney stones, etc.

In the present paper, we concentrate on the modeling aspects of such an approach, outlining possible minimum requirements for phase-field modeling of biological crystallization processes, and demonstrate that with appropriate choice of the model parameters and boundary conditions phase-field models can approximate the polycrystalline microstructure formed in simple cases of biomineralization (shell formation in mollusks such as bivalves, gastropods and cephalopods, and sprinkle formation in coral skeletons).

## 2. Microstructures formed during biomineralization

Before outlining the phase-field models we used in the present research, we give a short account of the experimental results on the observed microstructures. Polycrystalline microstructures formed in biomineralization processes have been investigated by a variety of experimental methods, including optical microscopy (OM), scanning electron microscopy (SEM), electron back scattering diffraction (EBSD), x-ray tomography, and polarization-dependent imaging contrast (PIC, [29, 30]), etc. Here we concentrate on experimental results obtained on the microstructure of molluscan shells and coral skeletons.

### 2.1. Mollusk shells

Mollusk shells are complex organo-mineral biocomposites with a broad range of species dependent microstructures [13, 14, 30–39]. A schematic view of bivalve anatomy having a nacre-prismatic shell is shown in Fig. 1. Moving inwards, the sequence of the individual layers is as follows: the organic periostracum, a leathery "skin", that encloses the domain where biominerlization takes place, a dominantly mineral (calcium carbonate, CC) prismatic layer, and the nacre composed of CC tablets and organic interlamellar membranes, the submicron thick extrapallial liquid [31, 34–36], and then the outer calcifying epithelum layer of the mantle. Images showing typical microstructures of mollusk shells of similar type are displayed in Fig. 2 [13, 26, 37, 38].

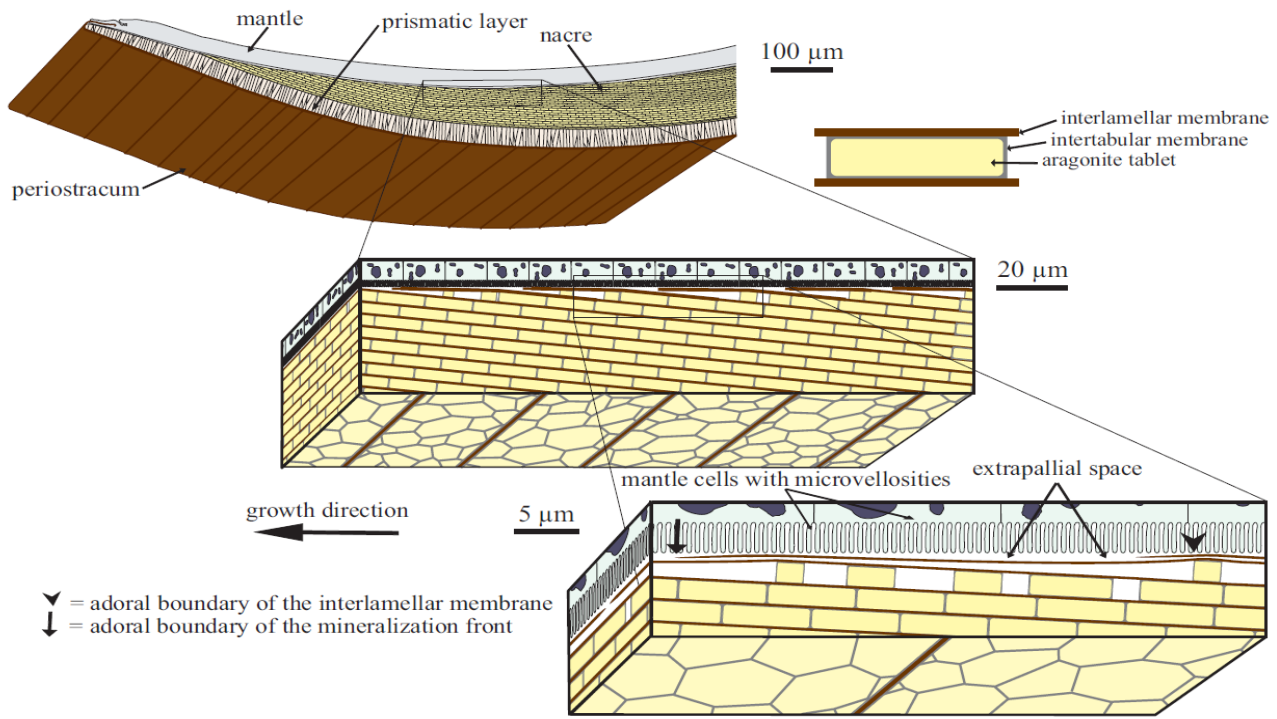

**Fig. 1.** Schematic view of bivalve molluscan anatomy with successive magnification of the mantle–nacre interface [33]. A thin liquid-filled extrapallial space is indicated (its thickness and content is open to debate). The interlamellar membrane is made of a viscoelastic chitin-based organic substance, whereas the mineral constituent is crystalline CC (aragonite). (Reproduced with permission from Ref. [33].)

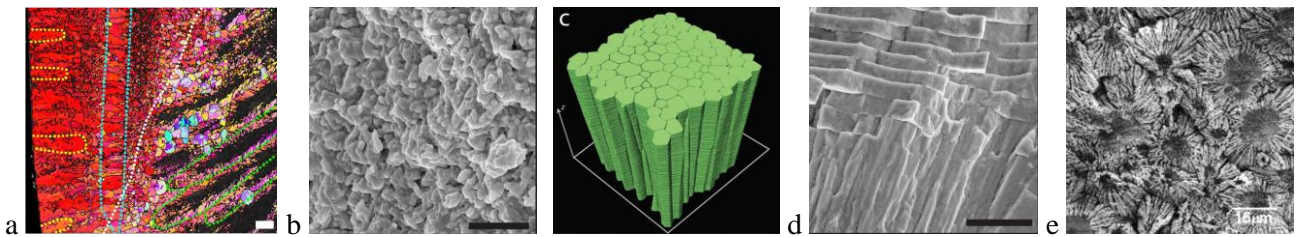

**Fig. 2.** Some typical mircrostructures observed in mollusk shells: (a) cross-sectional EBSD orientation map for the shell of *Katelysia rhytiphora* [37], the outer side is on the left (note the transitions between layers of different crystallite morphologies); (b)–(d) microstructures (SEM) observed by SEM: (b) outer randomly oriented granular domain in the shell of *Unio pictorum* [26]; (c) columnar prismatic domain of *Pinna nobilis* (x-ray tomography reconstruction) [13], and (d) plate-like structure of the nacre of *Unio pictorum* (SEM) [26]; (e) spherulitic layer, section perpendicular to growth in the shell of *Haliotis rufescens* [38]. (Reproduced with permission from Refs. [13, 26, 37, 38].)

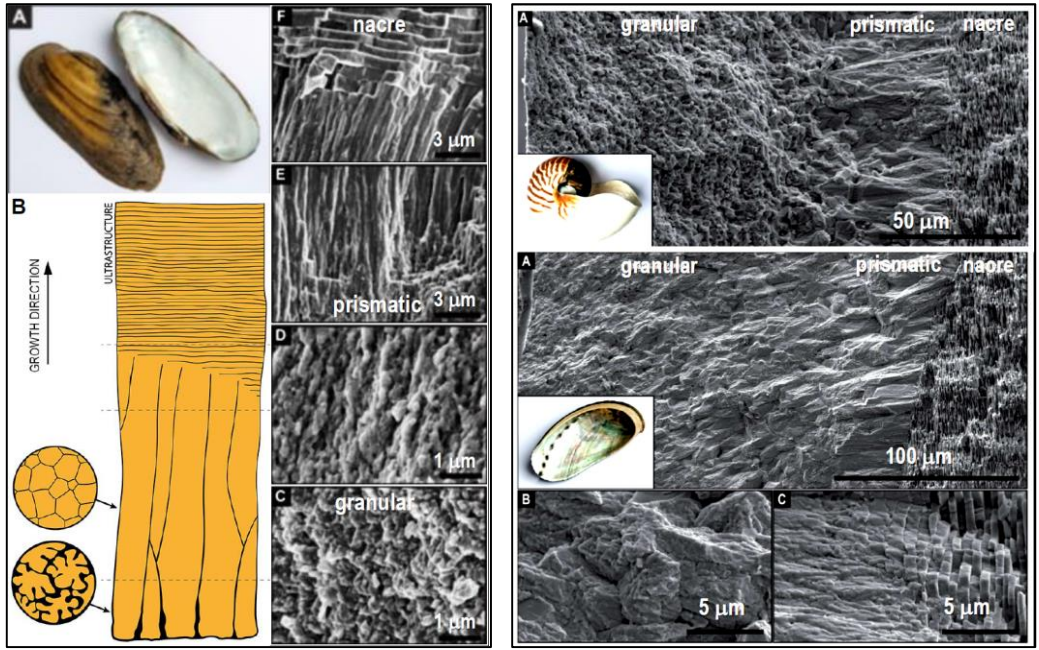

**Fig. 3** Hierarchy of ultrastructures in the shell of some of the mollusks. Left: for bivalve *Unio pictorum* (growth direction: upwards); A – The shell, B – shematic drawing of microstructure, C – F scanning electronmicroscopy (SEM) images. Right top: for cephalopod *Nautilus pompilius* (growth direction: to the right). Right bottom: for gastropod *Haliotis asinina* (growth direction: to the right); A-C – SEM image, C and D are magnified views of the granular domain and the prismatic → nacre transition. Note the similar sequence of ultrastructures during growth: granular → prismatic (columnar) → nacre (alternating mineral and organic layers). (Reproduced with permission from Ref. [27].)

A recent study shows that in members of three classes of mollusks *Unio pictorum* (bivalve), *Nautilus pompilius* (cephalopod), and *Haliotis asinine* (gastropod), the shell displays a common sequence of ultrastructures: a granular domain composed of randomly oriented crystallites, a prismatic domain of columnar crystallites, and the nacre [26, 27] (Fig. 3). It has been shown that the layered structure of nacre may contain screw dislocation-like defects (see Fig. 4) [14, 39, 40].

Of these structures the prismatic layers show mechanical flexibility, whereas the nacre (also called "mother of pearl") is fairly rigid but hard; the combination of the two yields a surprisingly strong yet flexible biocomposite (see e.g., Ref. 22).

### 2.2. Microstructure of the coral skeletons

The multiply branched shapes of coral skeletons are covered by a large number of coral polyps [41] and the connecting living tissue, which secretes calcium carbonate to create a hard shelter (the *corallite*, a tubular hollow structure on which the polyp sits) [42], into which the polyp can retreat if danger is detected (see Figs. 5a and 5b). The polyps are transparent, their color originates from photosynthesizing algae (*zooxanthallea*) that live in symbiosis with the polyp and feed the polyp sugars and oxygen. The surface of the skeleton is intricately structured [42], depressions, ridges, cavities are arranged into complex

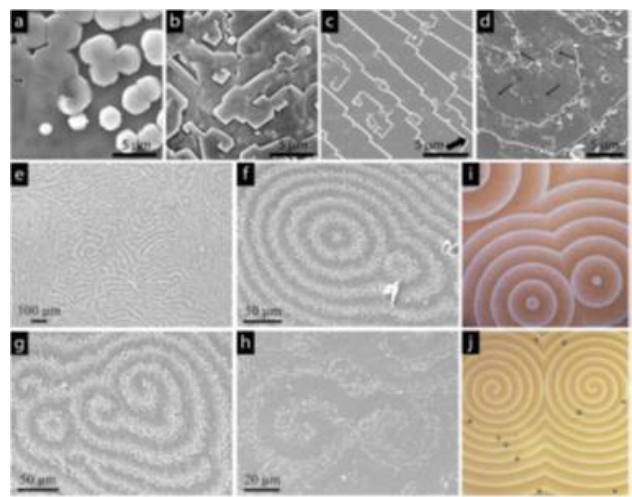

**Fig. 4** (a)–(h) Screw dislocation-like defects at the growth front of the nacreous layer of various species [14, 39]. For comparison (i), (j) images for target and spiral patterns formed by Belousov-Zhabotinsky reaction are also shown [40].

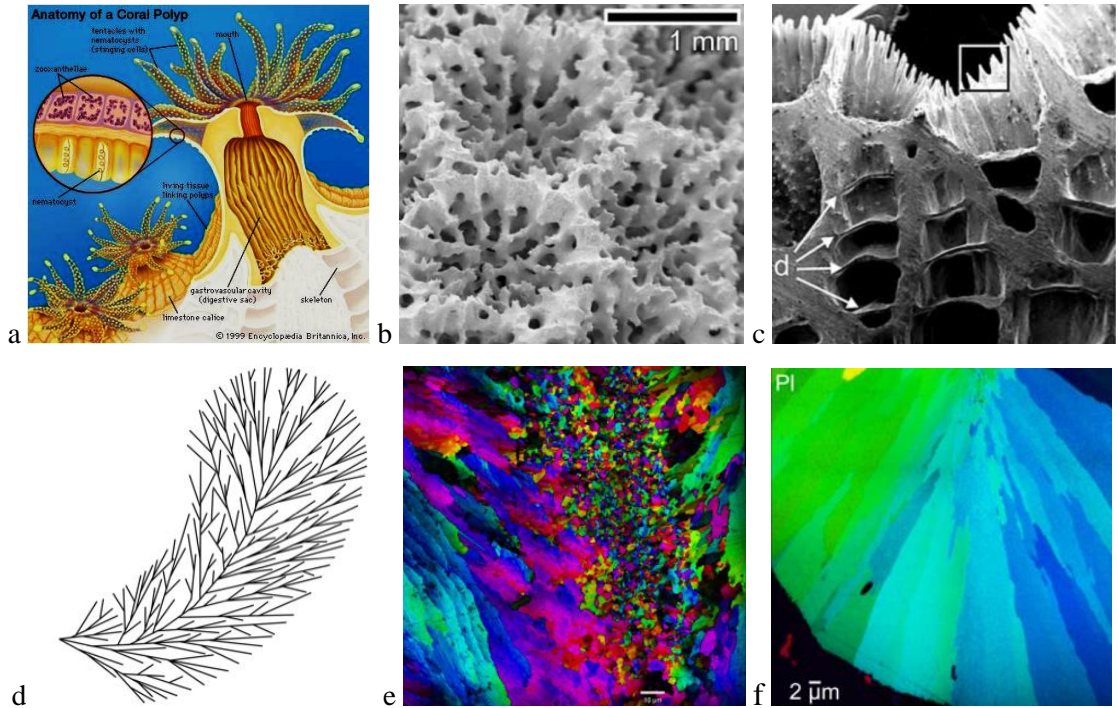

**Fig. 5.** From coral polyps to the CC (aragonite) skeleton they form, highly structured from the cm to the nm scale. (a) Schematic drawing of the coral polyp (colored) sitting on the porous skeleton (white) [41]; (b) SEM image of the surface formed below coral polyps (corallites) in the case of *Porites sp.* [42]; (c) cross-section of corallite, *Pollicipora damicornis* [43] (linear size of image is ~ 3 mm, "d" stands for mineral bridges termed distally convex dissepiments); (d) schematic drawing of a "plumose" spherulite [15]; (e) PIC map of *Acropora pharaonis* coral skeleton showing spherulitic microstructure of crystals radiating from a band of randomly oriented sprinkles (linear center of calcification, the bar indicates 10 μm) [28]; (f) PIC map of *Phyllangia* coral skeleton showing spherulitic microstructure of crystals radiating from centers of calcification, however, with no sprinkles [28].

patterns reflecting the radial symmetry of the polyp (Fig. 5b). The CC crystal (aragonite) building the porous skeleton has a spherulitic microstructure, that is, a radial arrangement of crystals radiating from a common center. The center does not have to be a point, it can be a line, or a plane in plumose spherulites. In the case of coral skeletons, the centers are curved planes, termed centers of calcification (CoCs) (Fig. 5d) [43, 15]. From these CoCs, acicular fibers grow radially, then arrange into fan-like bundles that finally group into a feather-duster-like shape termed "trabecula" [15]. Besides the plumose spherulitic structure, randomly oriented nanoscale crystallites "sprinkles" are also present [28, 44, 45] (Fig. 5e). Recent PIC mapping experiments performed at synchrotron indicate that the amount of the submicron sized sprinkle crystallites varies considerably among different coral species: in some of them the sprinkles are missing, whereas in others submicron size crystallites appear at the perimeter of the skeleton, along grain boundaries, at the growth front of spherulitic trabeculae, or in bands formed in the interior of the skeleton [28]. The discovery of sprinkles inspired a refined model for spherulitic growth in corals, described in detail in [28]. Randomly oriented sprinkles are the first nucleated crystals at the growth front. With further growth, those oriented radially have space and thus continue to grow, those oriented tangentially run into each other and stop growing. This is why they stay small. A coarsening process then makes the larger crystal grow larger at the expense of the smaller ones, which disappear. In most mature spherulites, therefore, no sprinkles remain. In the skeleton of some coral species, however, some sprinkles do not disappear, presumably because they are kinetically trapped.

In our previous work, we modeled this process by phase-field simulations [28] and raised the possibility that other spher-ulites may grow this way, including aspirin, chocolate, and geologic crystals. However, the formation mechanism of sprinkle bands and the origin of different amount of sprinkles in the skeleton of different coral species is not yet fully understood. Mathematical modeling is expected to help to identify the governing factors.

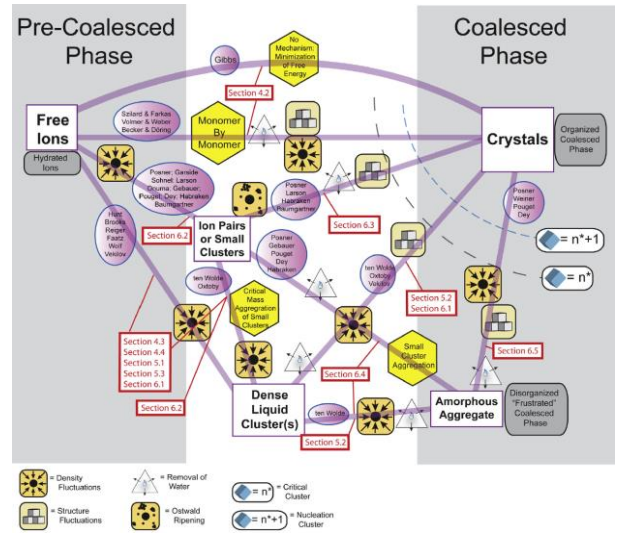

**Fig. 6.** Possible pathways of biomineralization as presented in Ref. [51]. The formation of a crystal is viewed as a discontinuous phase transition. The "classical" monomer-by-monomer pathway is represented by the uppermost straight line connecting the pre-coalesced free ions to the fully the coalesced crystals. "Non-classical" transition routes (see hexagons) proposed by authors listed in ellipses are also displayed. Steps (indicated by "road signs") required to make the transition from one state to the other (given in rectangles) are also indicated. The red section numbers refer to sections of Ref. [51], which describe these mechanisms in detail. In this work, we investigate two specific routes: (1) classical ion-by-ion addition and (2) the non-classical amorphous CC mediated crystallization. (Reproduced with permission from [51].)

*2.3. Biomineralization on the nano- and macro scale*

The complexity of biomineralization stems mainly from the fact that the fluid and solids incorporate organic molecules, the role of which is largely unknown [35, 36, 46, 47]. For example, nanoscale amorphous calcium carbonate (ACC) globules are essential for the formation of mollusk shells and coral skeletons [13, 14, 48–50]. A schematic drawing that shows possible pathways from free ions to crystalline CC are presented in Fig. 6 [51]. Evidently, all these processes cannot be explicitly included into an orientation field based phase-field (OF-PF) approach that models the polycrystalline microstructure on the mesoscale. In this work, we investigate two specific cases: formation of crystalline CC via the classical route of ion-by-ion addition of $Ca^{2+}$ and $CO_3^{2-}$, and (2) the non-classical route via an amorphous precursor (ACC) deposited at the solidification front. In the latter case, we hypothesize that crystallization rate is determined by the velocity of crystal growth into the ACC layer and not by the rate of supplementing ACC; i.e., growth is controlled by the self-diffusion in ACC. This hypothesis can account for the typically months' time scale of shell/skeleton formation, which would be difficult to interpret, e.g., on the basis of ion-by-ion deposition directly from the extrapallial fluid or other aqueous solutions. A specific realization of mechanism (2) is presented in Ref. [52]. The extrapallial fluid contains various ions and organic molecules as shown by in *vivo studies* [35, 36]. Molecular / ionic mobility in the extrapallial fluid is expected to be orders of magnitude higher than in the solid. As a result, in the case of growth via ion-by-ion addition, either a slow supply of the ions or a kinetic barrier of ion deposition can keep the growth rate sufficiently low. This offers limitations to the possible mechanisms, as will be discussed later.

The shell of bivalves grows typically by 100 – 300 μm lunar-day increments [53, 54], corresponding to a growth rate of about $v \approx 4.2\times10^{-11}$–$1.3\times10^{-10}$ m/s, which decreases with age [53]. The thickening rate of the shell of *Tridacna deresa* was estimated to be $v \approx 1.6\times10^{-10}$–$4.9\times10^{-10}$ m/s in its early life, which decreases to $v \approx 3.2\times10^{-11}$–$2.3\times10^{-10}$ m/s in the later life [55]. Comparable growth rates were reported for freshwater gastropods $v \approx 9.5\times10^{-11}$ m/s [56].

## 3. Phase-field modeling of polycrystalline structures

There are two main categories of the phase-field (PF) models developed to address polycrystalline freezing: (i) the multi-phase-field (MPF) models that assign a separate phase field for every crystal grain [57-63], and (ii) the orientation-field based phase-field (OF-PF) approaches, in which the local crystallographic orientation is monitored by scalar (2D) [64–71] or quaternion/rotation matrix fields (3D) [71–78]. Recent developments in these areas were reviewed in Ref. [71], respectively. In a recent study [63], it was shown that most of the previous MPF models do not satisfy evident consistency criteria, and a consistent model was proposed. Both the MPF and OF-PF approaches have their advantages, yet complex polycrystalline growth forms (such as disordered dendrites, crystal sheaves, various types of spherulites, and fractal-like poly-crystalline aggregates, etc.) were so far modelled exclusively using the OF-PF approach.

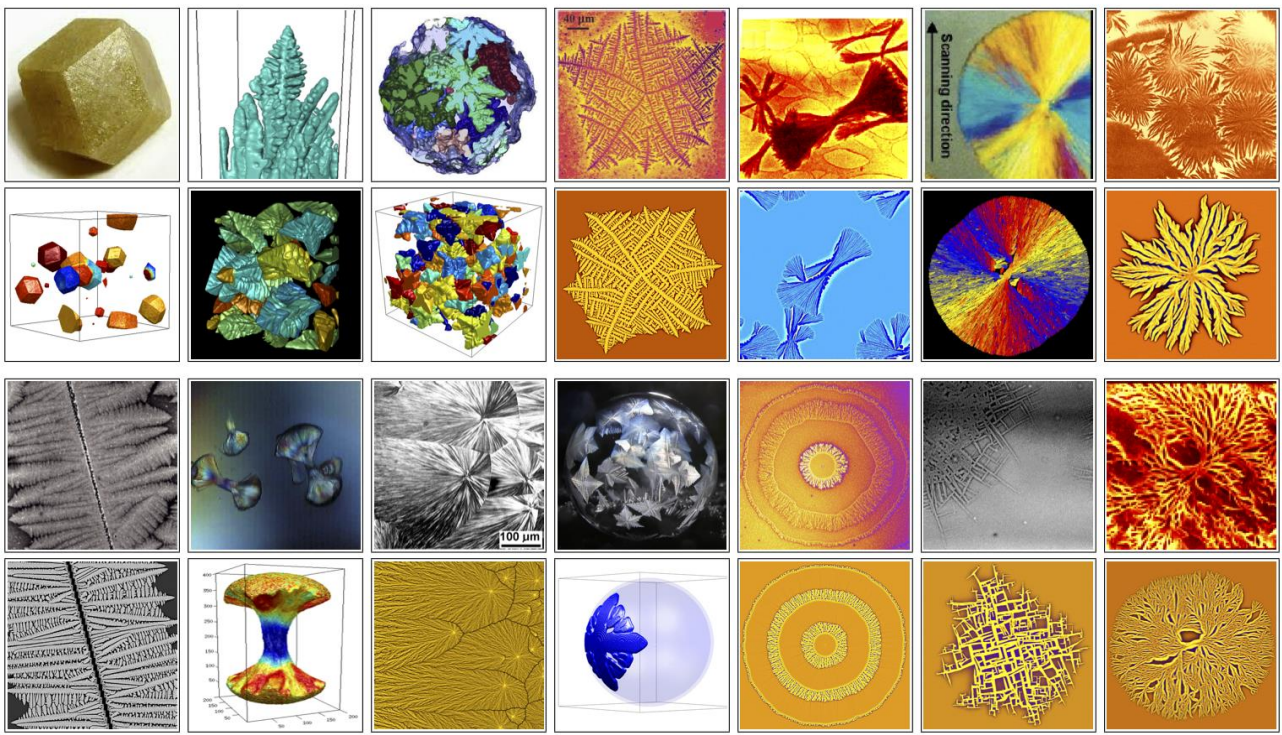

**Fig. 7.** Increasingly complex crystallization structures in the experiments (1st and 3rd rows) and in the corresponding simulations by the OF-PF models performed at the Wigner Research Centre for Physics (2nd and 4th rows). Upper block: From left to right: rhombic dodecahedron crystals [77], columnar dendrites [71], equiaxed dendrites [71], "dizzy" dendrites [67], and various types of spherulites (Refs. [69, 71]). Lower block: From left to right: scratch induced dendritic crystallization in polymer film [79], dumbbell-shape spherulites [71], spherulites in temperature gradient [71, 79], dendrite growing in a spherical shell, effect of oscillating temperature [71], "quadrites" forming by nearly 90° branching [69, 71], spherulitic structure formed from needle crystals [69, 71].

In the OF-PF models the local phase state of the matter is characterized by a coarse grained structural order parameter, the "phase field", that is time and space dependent, and monitors the transition between the liquid and solid states. This field is usually coupled to other slowly changing fields, such as the concentration field of the constituents, and the orientation field. The free energy contains bulk free energy and gradient terms for these fields. The equations of motion can be obtained via a variational route. The thermal fluctuations are represented by adding noise (non-conserved fields) or flux noise (conserved fields) to the equations of motion that satisfy the fluctuation-dissipation theorem, allowing for homogeneous nucleation [65] The inclusion of foreign surfaces / particles represented by appropriate boundary conditions that set the wetting properties, allows the modelling of heterogeneous nucleation [76–79] and solidification in confined space [79]. The addition of phase-field noise makes the modelling of Brownian motion of solid particles possible [80]. The OF-PF models may also be coupled to hydrodynamic flow, which can be represented by either the Navier-Stokes equation [81, 82] or the lattice Boltzmann technique. With an appropriate overlapping grid technique, flow coupled motion of growing solid particles can also be treated [83]. The main virtue of PF modeling is that the growth morphology can be computed on the basis of the freezing conditions and the thermophysical properties. Images obtained by OF-PF modelling in two- and three dimensions (2D and 3D) [67, 69, 71, 77, 79] are displayed in Fig. 7. Recently, PF modeling has been extended for microstructure formation in mollusk shells and coral skeletons, and promising agreement was seen between experiment and the predicted microstructures [26–28]. We give below a detailed account of the modeling efforts and present new results.

Herein, we continue further the quest for a *minimum phase-field model of the biomineralization process* during the formation of mollusk shells and coral skeletons. Evidently, we cannot model the living organism in the framework of this approach, their functions are represented by appropriate boundary conditions. Unfortunately, usually little is known of the thermodynamics of the multi-component fluids involved, of the interfacial free energies and their anisotropies, and the respective diffusion coefficients. Therefore, we aim at identifying the main components a minimal theory needs to incorporate for qualitatively reproducing the microstructures seen in the experiments.

In the present study, we rely on three specific formulations of the phase-field theory. In two of them [26, 27] the local state is characterized by the phase field, $\phi(\mathbf{r}, t)$ that monitors the process of crystallization, the concentration field $c(\mathbf{r}, t)$ specifying the local composition, and a scalar orientation field $\theta(\mathbf{r}, t)$, that represents the local crystallographic orientation in 2D. The latter field is made to fluctuate in time and space in the liquid, a feature that represents the short range order present in the liquid state (as done in Refs. [65, 67–71] and [76–79]). The third model does not include an orientation field; it has been developed to handle two-phase spiraling dendrites during eutectic solidification in ternary alloys in 3D [84] with fixed relative orientation of the solid phases.

### 3.1. Phase-field model 1 (PF1)

The first model will be used to address shell formation in the case of mollusks and the formation of coral skeletons. This model is similar to the standard binary PF theory by Warren and Boettinger [85], however, it is supplemented with an orientation field as done in Refs. [65, 67–71]. Accordingly, the OF-PF free energy of the heterogeneous system is expressed as

$$F = \int d\mathbf{r} \{ \frac{\varepsilon_\phi^2 s^2 T}{2} (\nabla \phi)^2 + w(c)Tg(\phi) + \\ + f_{chem}(\phi, c) + f_{ori}(\phi, \nabla \theta) \}. \quad (1)$$

Here, $T$ is the temperature. Parameters $\varepsilon_\phi^2 = (12/\sqrt{2})\, \gamma_i\, \delta_i\, /T_i$ and $w(c) = (1 - c)\, w_A + c\, w_B$, where $w_i = (12/\sqrt{2})\, \gamma_i\, /(\delta_i T_i)$ are expressed in terms of the free energy $\gamma_i$, the thickness $\delta_i$ of the crystal-liquid interface, and the melting point $T_i$ of the $i^{th}$ pure component ($i$ = A or B that stand for the organic and mineral components, respectively). $s = s(\vartheta, \theta) = 1 + s_0 \cos\{k\vartheta - 2\pi\theta\}$ is an anisotropy function corresponding to an interfacial free energy of $k$-fold symmetry and strength $s_0$, whereas $\vartheta = \arctan(\phi_y/\phi_x)$ is the angle of the normal of the interface in the laboratory frame, while $\nabla \phi = [\phi_x, \phi_y]$. The angular (circular) variables $\vartheta$ and $\theta$ are normalized so that they vary between 0 and 1. The bulk free energy density reads as,

$$f_{chem} = p(\phi) f_C(c, T) + [1 - p(\phi)] f_M(c, T), \quad (2)$$

and varies between the free energy densities of the crystal and mother phases ($f_C$ and $f_M$, respectively) as prescribed by the interpolation function $p(\phi) = \phi^3(10 - 15\phi + 6\phi^2)$. Here $f_C$ and $f_M$ were taken from the ideal solution model. The orientation free energy density is as follows:

$$f_{ori} = p(\phi) H |\nabla \theta|, \quad (3)$$

where the parameter $H$ can be used to tune the magnitude of the grain boundary energy.

The time evolution of the heterogeneous system is described by variational equations of motion (EOMs):

$$\frac{\partial \phi}{\partial t} = -M_\phi \frac{\delta F}{\delta \phi} + \zeta_\phi, \quad (4)$$

$$\frac{\partial c}{\partial t} = \nabla \cdot \left\{ M_c \nabla \frac{\delta F}{\delta c} + \zeta_c \right\}, \quad (5)$$

$$\frac{\partial \theta}{\partial t} = -M_\theta \frac{\delta F}{\delta \theta} + \zeta_\theta, \quad (6)$$

where $M_\phi$, $M_c$ and $M_\theta$ are mobilities that determine the time-scale of the evolution of the individual fields, and are related to coefficients of the self-diffusion, interdiffusion, and rotational diffusion [65, 67–71]. The chemical and orientation mobilities are made phase-dependent as $M_i = M_{i,C} + p(\phi)\{M_{i,M} - M_{i,C}\}$, where $i = c$ or $\theta$, and indices $M$ and $C$ denote values for the mother and crystalline phases. Gaussian white noise terms $\zeta_i$ are added to the EOMs to represent the thermal fluctuations (here $i = \phi$, $c$, and $\theta$). 3D generalizations of model PF1 can be found elsewhere [71–74]. A recent substantial extension of the OF models [75] removed limitations of previous OF-PF models, via incorporating tilt- and inclination dependencies besides the misorientation dependence of the grain boundary energy.

## 3.2. Phase-field model 2 (PF2)

The second model will be used to provide a refined model of mollusk shell formation, and the nacreous structures. This OF-PF model was originally developed to describe eutectic solidification, while keeping a fixed orientational relationship between the two solid phases inside the crystal grains [86]. To realize this, a square-gradient term, $\frac{1}{2}\varepsilon_c^2 T (\nabla c)^2$, was added to the free energy density, and a more complex orientational free energy term was incorporated,

$$f_{ori} = p(\phi)H\{h(c)F_1(|\nabla\theta|) + [1 - h(c)] F_2(|\nabla\theta|) + (\varepsilon_\theta^2 H/2T)|\nabla\theta|^2\} \quad (7)$$

that realizes a fixed orientational relationship at the solid-solid phase boundaries. Here $h(c) = \frac{1}{2}\{1 + \cos[2\pi (c - c_\alpha)/(c_\beta - c_\alpha)]\}$, $c_\alpha$ and $c_\beta$ are the CC concentrations in the two solid solution phases, whereas $F_1(|\nabla\theta|) = |\nabla\theta|$ and $F_2(|\nabla\theta|) = a + b|\cos(2m\pi d|\nabla\theta|)|$. Here $a$, $b$, $m$, are constants, and $d$ is the characteristic thickness of the sold-solid phase boundary. The EOMs are derived the same way as in the case of model PF1.

## 3.3. Phase-field model 3 (PF3)

Using this model, we are going to address screw dislocation-like 3D topological defects observed in the nacre. The free energy of the solidifying system is given by the expression [84]

$$F[\phi, \mathbf{c}] = \int d\mathbf{r} \left\{ \frac{\varepsilon_\phi^2 T}{2}(\nabla\phi)^2 + wTg(\phi) + [1 - p(\phi)]f_M(\mathbf{c}) + p(\phi)\left[f_C(\mathbf{c}) + \frac{\varepsilon_c^2 T}{2}\sum_{i=1}^{3}(\nabla c_i)^2\right]\right\}, \quad (8)$$

where $\mathbf{c} = [c_1, c_2, c_3]$ and $\sum_i c_i = 1$, whereas $w$ and $\varepsilon_c^2$ are constants. The bulk free energies of the solid and liquid phases are taken from the regular and ideal solution model:

$$f_C(\mathbf{c}) = \sum_{i=1}^{3} c_i[f_i^C + \log c_i] + \frac{1}{2}\sum_{i,j,i\neq j}^{3} \Omega_{i,j} c_i c_j \quad (9)$$

and

$$f_M(\mathbf{c}) = \sum_{i=1}^{3} c_i[f_i^M + \log c_i], \quad (10)$$

where $\Omega_{i,j}$ are the binary interaction coefficients in the solid.

The respective EOMs obtained variationally are as follows

$$\frac{\partial \phi}{\partial t} = s(\mathbf{n})\left\{\varepsilon_\phi^2 \nabla^2\phi - wg'(\phi) + p'(\phi)[f_M(\mathbf{c}) - f_C(\mathbf{c})] - p'(\phi)\frac{\varepsilon_c^2}{2}\sum_{i=1}^{3}(\nabla c_i)^2\right\}, \quad (11)$$

and

$$\frac{\partial c_i}{\partial t} = \sum_{j=1}^{3} \nabla \cdot \left[(1 - p(\phi))M_{i,j}^c\left(\nabla \frac{\delta F}{\delta c_j}\right)\right], \quad (12)$$

where $\mathbf{M}^c$ is the 3 × 3 mobility matrix. With the choice of 1 and −0.5 for the diagonal and off-diagonal elements the criterion $\sum_i c_i = 1$ is automatically satisfied. The diffusion is switched off in the bulk solid. Further information on the model PF3 is available in Ref. [84].

## 3.4. Numerical solutions

The EOMs were solved numerically in a dimensionless form, on rectangular uniform grids, using finite difference discretization with forward Euler time stepping. The PF1 and PF2 codes were run on a CPU cluster of 320 CPU cores using MPI protocol. Typical runs on a 1000×2000 grid took between about 8 to 15 hours, depending on the number of time steps that varied from 2.4×10⁵ to 5×10⁵, as required by the velocity of crystallization. The code for PF3 was run on high end graphics processing units (GPUs), and was solved on a 3D rectangular grid.

## 4. Materials parameters

Since in the biomineralization problems we address here the dominant CC polymorph is the metastable aragonite, we use the thermophysical data available for this polymorph, as much as possible. In case, where no information is available, we use values for another CC polymorph (calcite).

The OF-PF models require a fairly detailed information on the systems studied. This incorporates the free energy of all the relevant phases as a function of temperature and composition, all the interface energies, and the translational-, chemical-, and rotational diffusion coefficients. Unfortunately, only limited thermophysical information is available even for the pure water-CC system from experiment and MD simulations, such as the phase diagrams [87, 88] and the equilibrium shapes reflecting the anisotropy of the interface energy [89, 90]. During biomineralization, however, a variety of ions and organic macromolecules are present that may influence/control the crystallization process [91–97]. Accordingly, it is a nontrivial task to obtain accurate input data for mesoscale modelling; as e.g., selective adsorption of ions or organic molecules on different crystal faces may change growth morphology [93, 94], or influence the formation of polymorphs of CC [96, 97].

Owing to this lack of information, we present here generic approaches, based on simplified hypothetical model systems of properties similar to those used in Refs. [26–28]. The input data are collected in Tables 1 and 2 for Model PF1, in Tables 3 and 4 for Model PF2, a few common ones are presented in Table 5.

We address here two scenarios for the formation of crystalline CC (CCC): (i) diffusion controlled growth of crystalline CC directly from aqueous solution via ion-by-ion addition; (ii) diffusion controlled growth of crystalline CC into a hydrated ACC (hACC) layer that is assumed to be deposited on the solidification front (by vesicles or the-ion-by-ion process) with a sufficient rate so that deposition is not the rate limiting process. These scenarios differ in the diffusion coefficient we assign to the "mother phase" (aqueous solution of ACC) that crystallizes. Since the equations of motion were made dimensionless using the diffusion coefficient of the mother phase, and we assume that the relative magnitudes of $M_\phi$, $M_c$, and $M_\theta$ remain the same in the mother phase independently whether it is liquid or amorphous, the two scenarios differ in only the dimensionless mobilities assigned to the crystalline phase. In aqueous solutions the coefficient of ion diffusion at room temperature is in the order of $D_L \approx 10^{-9}$ m²/s [98]. The ion diffusion in ACC at 300 K is in the order of $D_{ACC,ion} \approx 10^{-15}$ m²/s [99]. Whereas the diffusion coefficient of the water molecules from molecular dynamics simulations is typically $D_{hACC,H2O} \approx 10^{-14}$–$10^{-13}$ m²/s for the slow H₂O molecules, although a few percent of H₂O molecules that have orders of magnitude faster diffusion ($D_{hACC,H2O} \approx 10^{-11}$ m²/s) are also present [99, 100]. However, biogenic ACC is almost anhydrous [101]. Therefore, water diffusion is expected to play a negligible role. The rate limiting factor for the structural transition is expected to be the slowest of these processes; accordingly, we use the diffusion coefficient for the ions, $D_{ACC,ion} \approx 10^{-15}$ m²/s [99]. We are unaware of self-diffusion data for aragonite. There are, however, experimental data for calcite. The Mg diffusion coefficient in calcite is about $D_{calcite} \approx 10^{-21}$ m$^2$/s at 823 K, which, extrapolates to $D_{calcite} \approx 10^{-53}$ m$^2$/s at room temperature provided that the diffusion mechanism does not change [102]. A different estimate is obtained via extrapolating the Mg-Ca interdiffusion data of Ref. [103] to room temperature, which yields $D_{calcite} \approx 10^{-29}$ m$^2$/s at 300 K for Mg diffusion in calcite. Even higher values emerge from MD simulations: $D_{calcite} \approx 10^{-22}$–$10^{-21}$ m$^2$/s [99], which may imply that the MD data somewhat overestimate the diffusion coefficient in the condensed phases. Herein, we opt for $D_{ACC,ion} \approx 10^{-16}$ m$^2$/s and $D_{calcite} \approx 10^{-29}$ m$^2$/s for ion diffusion in ACC and calcite, respectively; assuming thus that the diffusion data for calcite can be viewed as a reasonable order-of-magnitude estimate for the other polymorphs, including aragonite.

The corresponding dimensionless mobility data, $m_\phi = M_\phi \varepsilon_\phi^2 T / D_{c,M}$, $m_c = M_c / D_{c,M}$, where $M_c = (v_m/RT)c(1-c)D_c$, and $m_\theta = M_\theta \xi HT / D_{c,M}$, are presented in Table 1. Here $v_m$ is the average molar volume of the components, $R$ the gas constant, $\xi$ the length scale, whereas $HT$ is the energy scale of the grain boundary energy.

Table 1. Dimensionless mobility coefficients for Model PF1.

|  | $m_\phi$ | $m_{c,M}$ | $m_{c,C}$ | $m_{\theta,M}$ | $m_{\theta,C}$ |
|---|---|---|---|---|---|
| aq. sol. → CCC | 3.75 | 1.0 | $10^{-20}$ | 120 | $120 \times 10^{-20}$ |
| ACC → CCC | 3.75 | 1.0 | $10^{-14}$ | 120 | $120 \times 10^{-14}$ |

Subscripts $M$ and $C$ stand for the mother and crystalline phases. The chemical mobility of the former was used as reference, as its chemical diffusion coefficient was used in making the EOMs dimensionless.

Table 2. Dimensionless thermodynamic data used in Model PF1 (ideal solution thermodynamics [26])

| Quantity |  | Value |
|---|---|---|
| $T_r$ | $= T/T_A$ | 0.911 |
| $T_{r,B}$ | $= T_B/T_A$ | 0.786 |
| $\Delta g_A$ | $= \Delta G_A/RT$ | $-0.1184$ |
| $\Delta g_B$ | $= \Delta G_B/RT$ | 0.1554 |

Here $\Delta G_{A,B} = \Delta H_{A,B}(T - T_{A,B})/T_{A,B}$ ($A$ stands for CC and $B$ for the organic component).

Table 3. Dimensionless mobility coefficients for Model PF2.

|  | $m_\phi$ | $m_{c,M}$ | $m_{c,C}$ | $m_{\theta,M}$ | $m_{\theta,C}$ |
|---|---|---|---|---|---|
| aq. sol. → CCC | 0.0144 | 1.0 | $10^{-20}$ | 12 | $12 \times 10^{-20}$ |
| ACC → CCC | 0.0144 | 1.0 | $10^{-14}$ | 12 | $12 \times 10^{-14}$ |

Subscripts $M$ and $C$ stand for the mother and crystalline phases. The chemical mobility of the former was used as reference, as its chemical diffusion coefficient was used in making the EOMs dimensionless.

Table 4. Dimensionless thermodynamic data used in Model PF2 (regular solution thermodynamics [27]).

| Quantity |  | Value |
|---|---|---|
| $T_r$ | $= T/T_E$ | 0.720 |
| $T_{r,A}$ | $= T_A/T_E$ | 1.169 |
| $T_{r,B}$ | $= T_B/T_E$ | 1.286 |
| $\Delta g_A$ | $= \Delta G_A/RT$ | $-0.5802$ |
| $\Delta g_B$ | $= \Delta G_B/RT$ | $-1.0477$ |
| $\omega_M$ | $= \Omega_M/RT$ | 2.0510 |
| $\omega_C$ | $= \Omega_C/RT$ | 3.6335 |

Here $\Omega_{M,C} = \Omega_{0,M,C} - \Omega_{1,M,C} T$.

Table 5. Materials and computational data used in PF1 and PF2.

| Quantity | Value | Unit | Ref. |
|---|---|---|---|
| $\gamma_A$ (CCC – aq. sol.) | 150 | mJ/m$^2$ | [100,102] |
| $\gamma_B$ (organic – aq. sol.) | 118 | mJ/m$^2$ | this work |
| $\gamma_A$ (CCC – ACC) | 87 | mJ/m$^2$ | this work |
| $\gamma_B$ (organic – ACC) | 68 | mJ/m$^2$ | this work |
| $v_m$ (CCC – aq. sol.) | 26.7 | cm$^3$/mol |  |
| $v_m$ (CCC – ACC) | 32.4 | cm$^3$/mol |  |
| $\xi$ | $2.1 \times 10^{-6}$ | m |  |
| $\delta$ | $4.15 \times 10^{-8}$ | m |  |
| $\Delta x$ | $6.25 \times 10^{-3}$ | - |  |
| $\Delta t$ | $4.75 \times 10^{-6}$ | - |  |

Experimental data for the water-aragonite interfacial free energy cover the range of 83 – 150 mJ/m$^2$ [104–107]. A recent *ab initio* theoretical treatment provides a considerably larger value (280 mJ/m$^2$), and information on its anisotropy for small aragonite clusters [108]. Herein, we use 150 mJ/m$^2$. We are unaware of data for the free energy of the ACC-aragonite interface. Using Turnbull's relationship for the interfacial free energy, $\gamma = \alpha \Delta H / (N_0 v_{mc}^2)^{1/3}$, where $\alpha$ is a constant, $\Delta H$ heat of transformation, $N_0$ the Avogadro-number, and $v_{mc}$ is the molar volume of the crystalline phase, a crude estimate can be made on the basis of the enthalpy difference between ACC and aragonite: $\Delta H_{calcite\text{-}ACC} = (14.3 \pm 1.0)$ kJ/mol [101]. A similar value may be expected for aragonite [101]. Considering $\alpha = 0.55$ from molecular dynamics simulations [109], one obtains $\gamma_{aragonite\text{-}ACC} \approx 87$ mJ/m$^2$, which result, however, needs independent confirmation by other experimental/theoretical methods. Once the thermodynamic data are fixed for components A and B, and the interfacial free energy is given for one of the components, Models PF1 and PF2 predict the interfacial free energy for the other component, provided that the interface thicknesses are similar [85]. For materials of comparable entropy of transformation this realizes $\gamma \propto T_{tr}$, where $T_{tr}$ is the temperature of the phase transition, a relationship that works well for the solid-liquid interfacial free energy of metals [110].

It is appropriate to emphasize here that owing to uncertainties of the thermodynamic diving force of crystallization and of the interface energy estimates, the simulations we present in the next section can only be regarded qualitative; they are aimed at demonstrating that phase-field modeling is able to capture various morphological aspects of biomineralization.

## 5. Results and discussion

### 5.1. Modeling of microstructures mimicking mollusk shells

*5.1.1 Shell-like microstructure in Model PF1*

In a recent OF-PF study [26], we made the following assumptions, when modeling the formation of mollusk shells within Model PF1:
- The CC crystals grow into the extrapallial fluid by the molecule/ion attachment mechanism.
- Binary ideal solution thermodynamics (CC and organic component) is applies. Evidently, treating the extrapallial fluid as a quasi-binary solution is a gross simplification. During crystallization CC-rich crystal, and an organic-component-rich "fluid" forms from the original homogeneous mixture. This construction was used as a simple means to provide thermodynamic driving force for CCC precipitation.

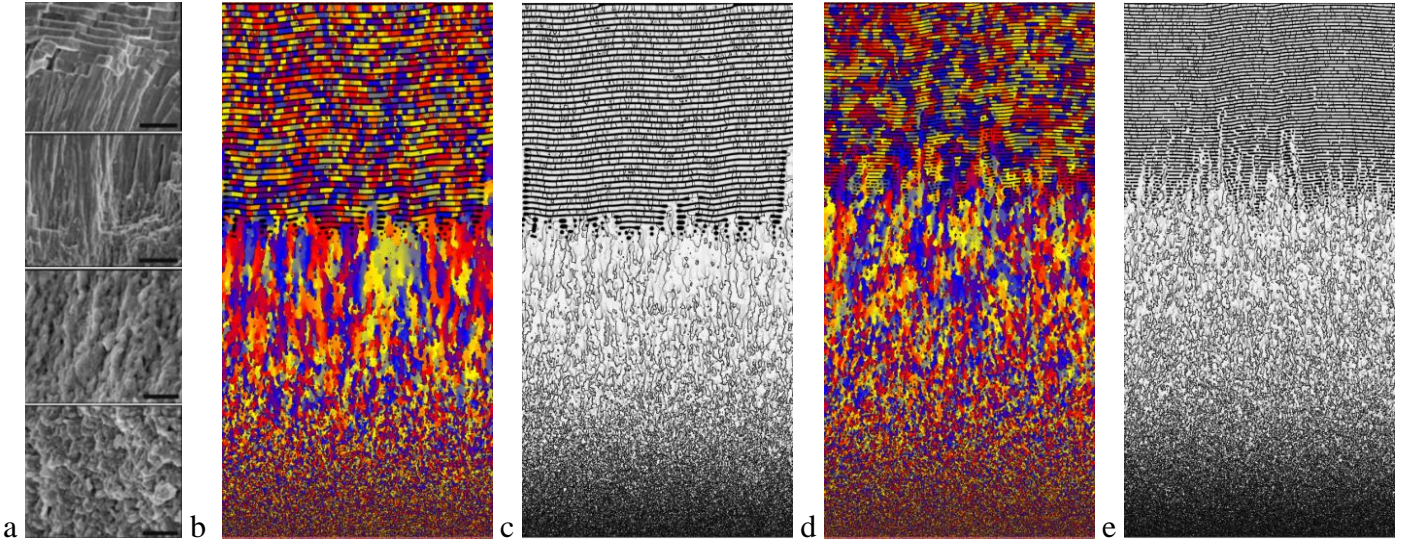

**Fig. 8.** Comparison of the microstructure of (a) the shell of mollusk *Unio pictorum* [24] (see also Fig. 3(c)) as shown by electron microscopy images with simulations (b)–(e) obtained by Model PF1 assuming (b), (c) *ion-by-ion attachment from extrapallial fluid* ($D_{c,M} = 10^{-9}$ m$^2$/s) or (d), (e) *crystallization from an ACC layer* ($D_{c,M} = 10^{-15}$ m$^2$/s). Growth direction is upwards. Orientation (b), (d) and grain boundary (c), (e) maps are shown. In the experimental images of (a), the bars correspond to 3 μm, 3μm, 1 μm, and 1 μm, respectively from top to bottom. Note the presence of the three characteristic domains in the experiment and in both types of simulations: granular, columnar prismatic, and sheet nacre structures. The simulations were performed on 1000×2000 grids. (In the high driving force (lower) part of both these simulations $l_D v / D_{c,M} > 1$; i.e., diffusionless crystallization takes place, whereas in the upper domain alternating mineral and organic layers mimicking the sheet nacre form.)

- CC-supersaturation of the extrapallial fluid decreases exponentially with the distance $x$ from the periostracum, owing to a spatially dependent amount of the organic component: $c(x) = c_{min} + (c_{max} - c_{min})\{1 - \exp(-9x/L)\}$, where $L$ is the thickness of the extrapallial space.
- Crystallization of CC starts via heterogeneous nucleation on the periostracum.
- The anisotropy of the CCC-mother phase interfacial free energy is neglected.

In the present work, besides this, we explore a different scenario shown in Fig. 6, in which the CCC crystal grows into an ACC precursor that forms continuously ahead of the crystallization front. The CCC front propagates into this ACC layer, and has no direct contact with the extrapallial fluid. It is also assumed that the formation of the ACC layer is fast enough, so that it is not the rate limiting process. The respective amorphous → crystal transition has in principle a reduced driving force, while orders of magnitude smaller diffusion coefficients prevail in the amorphous phase, yielding a far longer time scale for crystallization, when compared to crystallization from an aqueous solution. In this scenario, the phase field monitors the amorphous → crystal transition, rather than the crystallization of a liquid. We furthermore assume that the coefficients of the translational, chemical, and rotational diffusion decrease proportionally by orders of magnitude during this process, retaining the same relative magnitudes of the mobilities as in the liquid (see Table 1). Since the EOMs are solved in dimensionless form, where time is de-dimensionalized using the chemical diffusion coefficient of the mother phase, the dimensionless chemical mobilities of the mother phase remain unchanged. What differs between the present computation and the previous one in [26] is the magnitude of the individual mobilities in the crystalline and the mother phase (see Table 1). Following the general principles of phase-field modeling (see e.g. Ref. [111]), the phase-field mobility is assumed to be independent of the phase. In contrast, in agreement with the experimental diffusion coefficients, the chemical and orientational mobilities are assumed to be 20 orders of magnitude larger in the extrapallial fluid then in the crystal, and 14 orders of magnitude larger in the amorphous CC phase than in the crystalline phase.

We, retain the assumptions made in Ref. [26] as listed above, however, with the difference that in the non-classical mechanism it is the CC content of the ACC layer that decreases exponentially (while the organic content increases) with the distance from the periostracum. For the sake of simplicity, we employ the same dimensionless driving force as in Ref. [26] for both the classical and the non-classical cases.

The microstructures evolved in the two cases are compared in Fig. 8. The characteristic microstructural transitions are present in both simulations. Whether from the fluid or the amorphous phase, first small randomly oriented CCC grains form in the neighborhood of the periostracum, of which crystal grains grow further inward yielding elongated crystals of random orientation that compete with each other. With increaseing distance from the periostracum, i.e., with decreasing supersaturation, growth slows down and the separation of the two constituents becomes possible, which leads to the formation of alternating CCC and organic-rich layers, closely resembling the nacre. The predicted sequence of the morphological transitions is similar for both mechanisms (i.e., for aq. sol. → CCC and ACC → CCC), however, there are minor differences in the relative thicknesses of the granular, columnar prismatic, and layered nacre structures. While the respective microstructures are rather similar, the typical size scales for the ACC → CCC transition is smaller than for the aq. sol. → CCC, roughly proportionally with the respective interfacial free energies. Alternating

CCC and organic-rich layers akin to sheet nacre form here due to a process described in Ref. [112], with the difference that under the present conditions a roughly flat "banded structure" forms, and thermal diffusion is replaced by chemical diffusion. Apparently, the orientational information is only partly transferred through the organic layers. Mineral bridges (discontinuities of the organic layers are also observed.)

The granular → columnar → layered morphological transitions occur here due to changes in the growth velocity. At high supersaturations nucleation and *diffusionless* solidification takes place forming the granular domain. At medium supersaturations nucleation ceases, only competing growth of the existing particles takes place yielding the columnar domain, whereas at small supersaturations alternating CCC and organic layers occur, forming a layered structure that closely resembles the sheet nacre.

Diffusionless crystallization is possible, when the diffusion length $l_D = D_{c,M}/v$ is comparable to the thickness of liquid/crystal or amorphous/crystal interfaces ($d \approx 10^{-9}$–$10^{-8}$ m [113, 114]), where $v$ is the growth rate. The transition from diffusion controlled to diffusionless growth takes place in the regimes $10^{-4} < dv/D_{c,M} < 1$ or $10^{-2} < dv/D_{c,M} < 10$, depending on the model [115]. Considering $l_D \approx 10^{-9}$ m and a typical experimental growth rate of $v \approx 10^{-10}$ m/s, one finds that the mother phase needs to have a diffusion coefficient of $D_{c,M} \approx 10^{-15}$–$10^{-20}$ m$^2$/s to show transition towards diffusionless growth on the time scale required. This clearly rules out the possibility that the CC crystals form dominantly by direct ion-by-ion attachment from the extrapallial fluid. In turn, this range of $D_{c,M}$ is consistent with crystallization from amorphous CC, as the magnitude of $D_{c,M}$ falls in the range diffusion coefficient takes in the amorphous state [116].

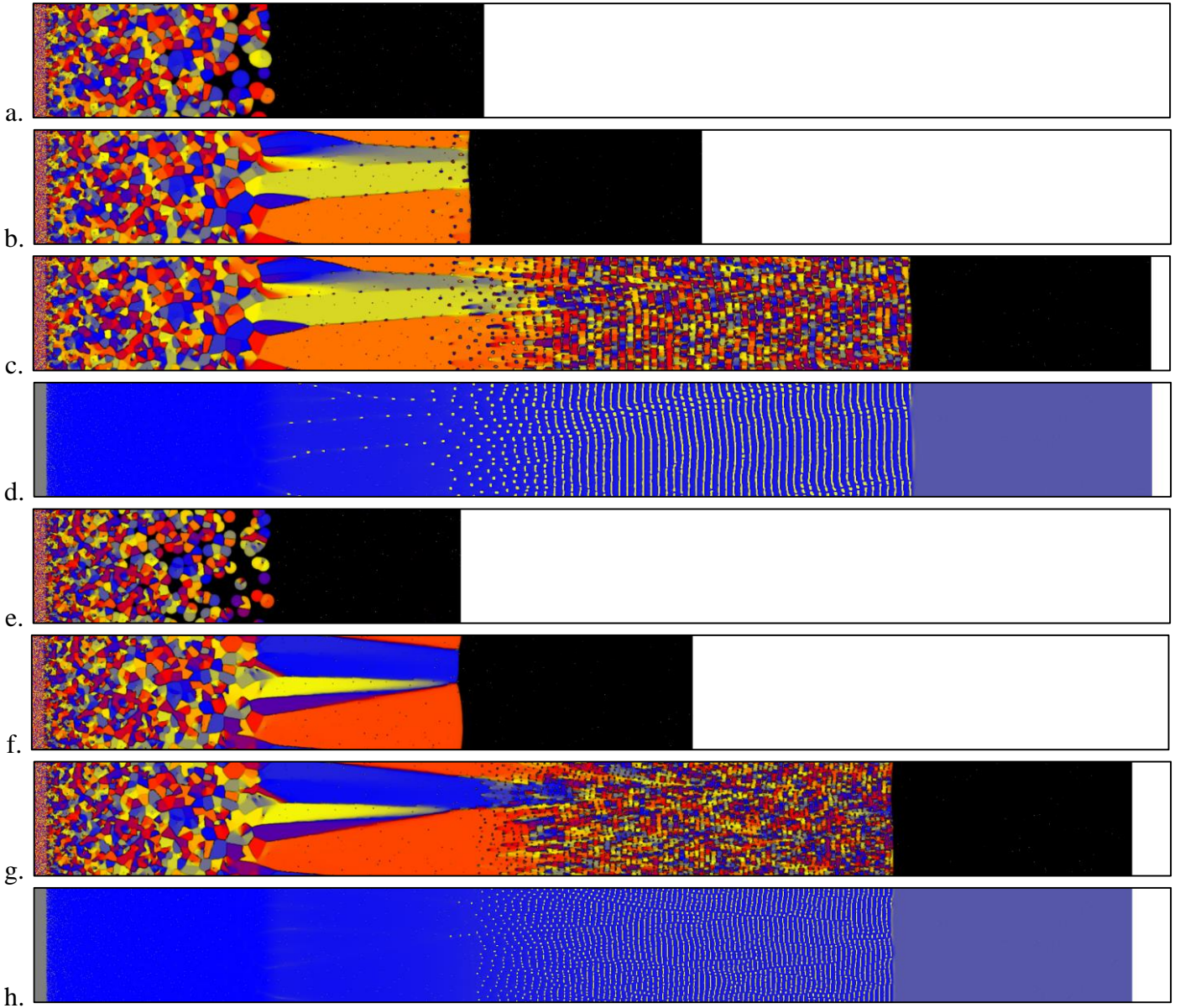

**Fig. 9.** Three stages of microstructure evolution in Model PF2 obtained assuming (a)-(d) *ion-by-ion attachment from extrapallial fluid* and (e)-(h) *by crystallization from ACC precursor* (orientation (a)-(c) and (e)-(g), and composition maps (d) and (h) are shown.): (a), (e) Formation of granular structure via heterogeneous nucleation dominated equiaxed solidification; (b), (f) columnar growth via directional solidification in concentration gradient yielding the prismatic structure, and (c), (g) layerwise formation of alternating CCC and organic layers (sheet nacre). In panels (a)-(c) the mother phase is an aqueous solution ($D_{c,M} = 10^{-9}$ m$^2$/s) in the simulation shown; whereas an amorphous precursor ($D_{c,M} = 10^{-15}$ m$^2$/s) is assumed in the simulation shown in panels (d)-(f). The wavelength of the layered structure is roughly proportional to the free energy of the mother phase – CCC interface. The thickness of the mother phase (extrapallial domain) is assumed to be constant. In (a)-(c) and (e)-(g) different colors stand for different crystallographic orientation, while colors white and black stand for the mantle of the mollusk and the mother phase, respectively. In (d) and (h) gray, yellow, and blue indicate the mother phase, the organic phase, and CCC. The simulations were performed on 2000×200 grids. In (a)-(c) and (e)-(g) time elapses downwards, whereas (c) and (d) and (g) and (h) displays snapshots taken at the same time.

Summarizing, while the two mechanisms considered here lead to similar microstructures, crystallization via the ACC precursor seems preferable to direct solidification via ion-by-ion addition from the extrapallial fluid, as in the latter case the diffusion coefficients are rather high, i.e., crystallization is expected to be fast, unless the "fluid" is of high viscosity (not realistic for the extrapallial fluid). Another problem of direct precipitation from the aqueous solution is that due to the high diffusion coefficient, the assumed initial exponential spatial dependence of CC supersaturation is only temporary on the time scale of shell growth, unless crystal growth is so fast that diffusional equilibration cannot take place. This is, however at odds with the experimental growth rates.

*5.1.2. Shell-like microstructures in Model PF2*

To relax some of the simplifying assumptions made in PF1, a refined model (PF2) was proposed for modelling the formation of mollusk shells in Ref. [27]. In this model, two solid phases form simultaneously from the liquid state, a mineral-rich and an organic-rich, while a fixed relative orientational relationship is forced between the solid phases formed inside the same crystal grain. This realizes a strong orientational coupling between the solid phases. Simultaneous formation of two solids occurs, e.g., in eutectic or peritectic systems. In the refined approach, we opted for the former case. The main assumptions made in deriving PF2 were as follows [27]:

- Crystal growth of CC happens via molecule/ion attachment.
- A binary eutectic model thermodynamics (regular solution) applies.
- Formation of granular CC crystals starts by heterogeneous nucleation on organic heterogeneities, whose number density is assumed to decrease exponentially with the distance from the periostracum.
- The thickness of the extrapallial domain (distance between the mantle and the solidification front) remains constant. (In the simulation, the position of the mantle surface varies in accord with the solidification rate.)
- The mineral content of the extrapallial fluid emitted at the surface of the mantle decreases exponentially with time.

In this approach, the assumption that CC supersaturation decreases towards the mantle is removed and is replaced by the more natural assumption that the CC supersaturation at the mantle decreases exponentially with time that leads to an analogous result, however, in a more natural way. The simulation presented in Ref. [27] shows a good qualitative agreement with the experimental microstructures observed for a various types of mollusk shells. It is reassuring that Model PF2 recovers the experimental microstructure though in a somewhat more ordered form, which could however be made more random by varying the noise added to the equations of motion. At the present state of affairs, it is difficult to decide which of models PF1 or PF2 should be considered superior to the other, yet the two-solid model is probably closer to the reality.

Herein, we explore whether the microstructure remains similar, when assuming ACC mediated crystallization in the framework of Model PF2. The results obtained by the two mechanisms of crystallization are compared Fig. 9. Apparently, as in the case of Model PF1, the two micromechanisms for crystalline CC formation yield similar microstructures.

Note that in Model PF2, the nacre is composed of two alternating solid phases. Under the conditions used in our simulation the organic component remains in an amorphous or nanocrystalline state. Remarkably, the predicted "nacre" structure recovers such details of the experimentally observed microstructure as the "mineral bridges" and the line defects across the organic layers termed "aligned holes" [117-121]. We will show, however, in the next section that despite this close similarity, the formation mechanism of the nacre can be considerably more complex than predicted here.

*5.1.3. Discussion of results from models PF1 and PF2*

First, we compare the solidification rates obtained from the simulations performed assuming $D_L \approx 10^{-9}$ m$^2$/s (aqueous solution) for the mother phase, with those using $D_{ACC,ion} \approx 10^{-15}$ m$^2$/s (ionic diffusion in ACC). We wish to stress that the velocities evaluated from the simulations can only be considered qualitative, as the thermodynamic driving force we used in this study might be well away from the true ones, which in turn may differ from the value obtained for the pure aqueous solution-CCC system, or the ACC-CCC system. Work is underway to perform more quantitative phase-field simulations to determine the growth and dissolution rates in pure systems, for which reasonably accurate experimental data are available [122].

The growth rate results are presented in Fig. 10. As one expects the average growth rate is roughly proportional to the diffusion coefficient of the mother phase. The growth velocities predicted for crystallization from ACC by models PF1 and PF2 are about two and one orders of magnitude higher ($v \approx 10^{-8}$ and $10^{-9}$ m/s, respectively) than the experimental ones, whereas the rates predicted for the ion-by-ion addition are about $v \approx 10^{-1}$ and $10^{-2}$ m/s, that are about 8-9 orders of magnitude too high. On this ground the mechanism based on the fast ion-by-ion addition can be excluded. One may perhaps argue that the production of calcium and carbonate ions in the surface layer of the mantle (outer epithelium) may be the rate limiting process, which may be then taken so slow as to match the experimental growth rate. However, in that case it is not the diffusion in the mother phase that controls the time scale of the process, and thus the mechanisms models PF1 and PF2 rely on would not be present. The position vs time relationships show characteristic differences for the two models: PF1 predicts a steeply increasing velocity, followed by a plateau decreasing slowly with time for the domain of alternating solid-liquid layers. In the present simulations the early stage behavior of the two models is different: while model PF1 starts with surface induced heterogeneous nucleation on the periostracum, PF2 relies on volumetric heterogeneous nucleation on organic impurities (a feature that can also be incorporated into PF1). As a result, in model PF2, fast initial crystallization is observed during the formation of the granular layer via volumetric heterogeneous nucleation. This is followed by steady state growth (roughly constant growth velocity) in both the prismatic and the nacreous domains, due to the lack of long range diffusion during fast eutectic solidification. In contrast, in PF1 a continuously decelerating solidification is observed, which is combined with oscillating growth rate in the nacre. The predicted mechanism (volumetric heterogeneous nucleation in a thick layer at the periostracum) that creates protection for the mollusk fast in the early stage of shell formation is advantageous from the viewpoint of survival. We note that due to computational limitations the maximum thickness of the computational domain is about 26.4 μm both in the PF1 and PF2 simulations. However, analogous structures can be produced on a larger size scale via reducing the rate by which the driving force of crystallization decreases with position/time and with an extended initial domain filled with heterogeneous nucleation centers.

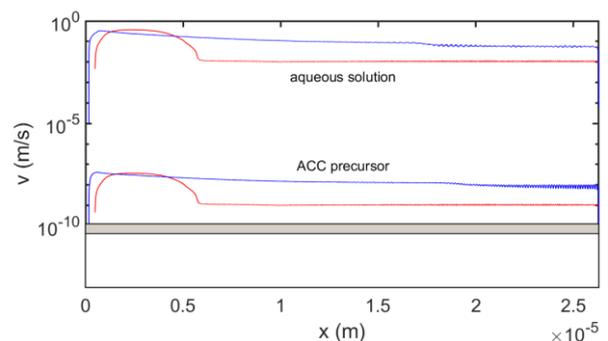

**Fig. 10.** Growth rates predicted by models PF1 (red lines) PF2 (blue lines) for the ion-by-ion mechanism from aqueous solution and for crystallization from ACC precursor. Note the ~ 7 orders of magnitude difference in growth velocity predicted for the two mechanisms. For comparison, the range of experimental data is also shown (grey domain). Note the oscillating growth rate in the layered domain.

It is appropriate to mention, that while our models describe the formation of the granular and prismatic layers, and the sheet nacre reasonably well within the framework of directional solidification, the predicted mechanism for the formation of the nacre via alternately precipitating mineral and organic layers may be oversimplified. This is especially true for columnar nacre: Experiments indicate that the formation of a quasi-periodic network of organic membranes precedes the formation of the CCC layers, which fill the space between the organic membranes later, as illustrated in Fig. 11 [117–121]. Although one could imagine that the organic membranes form periodically in space via an oscillating chemical reaction of the diffusion-reaction type, while the extrapallial fluid fills the space between the membranes, apparently, the real mechanism is more complex: first a multilayer outer membrane forms, of which the individual layers are exfoliated by nucleating CC crystals [117, 118].

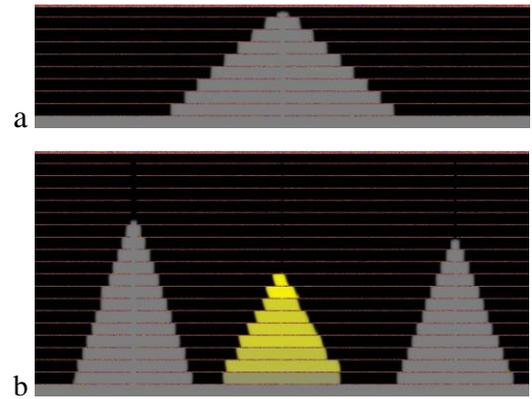

**Fig. 12.** Snapshots of 2D phase-field simulations (model PF1), showing the formation of columnar nacre-like CCC tile stacks (grey/yellow - different colors denote different crystallographic orientations) via solidification between preexisting organic membranes (brown horizontal lines) with mineral bridges provided by (a) aligned holes of uniform width (20 $\Delta x$), and (b) aligned holes of three different widths (from left to right, 20, 5, and 10 $\Delta x$, respectively). The simulations were performed on 2000 × 500 and 2000 × 1000 grids (a) with anisotropic interfacial free energy and with (b) kinetic and interfacial free energy anisotropy.

Summarizing, while models PF1 and PF2 cannot capture all details of the formation mechanism of nacre, remarkably similar microstructures are generated. This raises the possibility that on the basis of the mechanisms these models realize (diffusion controlled solidification at high driving forces), one may design and prepare artificial mollusk shell structures that inherit the mechanical excellence emerging from the hierarchical sequence of the granular, prismatic, and nacre-like ultrastructures. Thus, the present work opens up the way towards a novel design strategy for creating biomimetic / bioinspired composite materials.

Finally, to address the evolution of columnar nacre within the phase-field theory, one can incorporate preexisting organic membranes into the simulations "by hand", including the aligned holes seen in the experiments [117,118]. We used PF1 to model the formation CCC stacks shown in Fig. 11. The thin organic walls were assumed to have amorphous structure (local orientation varied randomly pixelwise) and we used a boundary condition that ensured a contact angle of 100° with the solid-liquid interface. Simulations of this kind yield 2D "pyramid-like" stacks of CC layers as shown in Fig. 12. Such simulations can be used to explore the effect of such parameters as growth anisotropy, contact angle, hole size/position, etc. For example, Fig. 12b implies that the growth velocity of stack height increases with increasing hole width.

*5.2 Helical structures predicted by model PF3*

Spectacular screw dislocation-like helical structures have been observed in mollusk shells akin to patterns formed in oscillating chemical reactions (Fig. 4). These 3D structures cannot be addressed in models PF1 and PF2 as the orientation field in them is valid in only 2D. Therefore, we use model PF3 to explore the possibility of forming such objects within the framework of the phase-field theory. Here two solid solution phases (α and β) precipitate from a homogeneous ternary liquid. Owing to the lack of relevant information, we retain the materials parameters used in Ref. [84]. Under appropriate conditions, shown in Fig. 13, a layerwise structure composed of alternating α (mineral) and β (organic) layers form, an analogue of the "nacre" observed in model PF2. In this regime, we see the formation of helical structures that emerge in pairs of opposite chirality. Different views and sections of such structures are shown in Fig. 14. This chiral structure closely resembles the screw dislocation-like defects reported in experiments [14, 39]. Apparently, our dislocation pairs do not recombine even in a long simulation time, presumably because of the lack of mechanical stresses that are not incorporated into the phase-field models used in this study. Further work is underway to characterize these structures and the dynamics of their formation in detail.

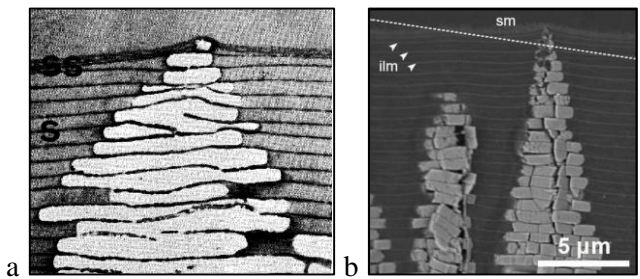

**Fig. 11.** Electron microscopy images displaying the formation of columnar nacre in the case of (a) *Haliotis rufescens* (reproduced with permission from [117]), and (b) *Phorcus turbinatus* (reproduced with permission from [118]). Here ss and s stand for "surface sheets" and "organic sheet", whereas sm and ilm indicate "surface membrane" and "interlamellar membrane", respectively.

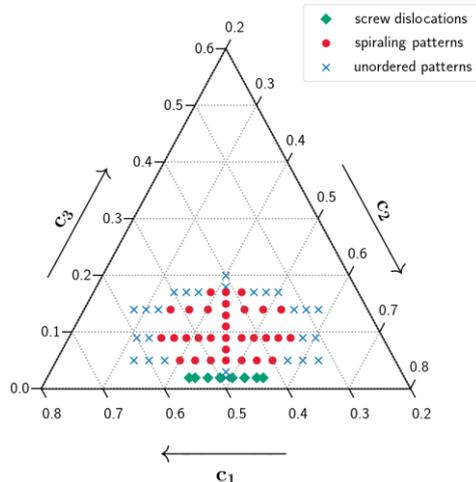

**Fig. 13.** Domain in an idealized ternary phase diagram [84], in which screw dislocation like structures form (green diamonds). For comparison domains of spiraling eutectic dendrites (red dots), and unordered eutectic structures (blue crosses) are also shown.

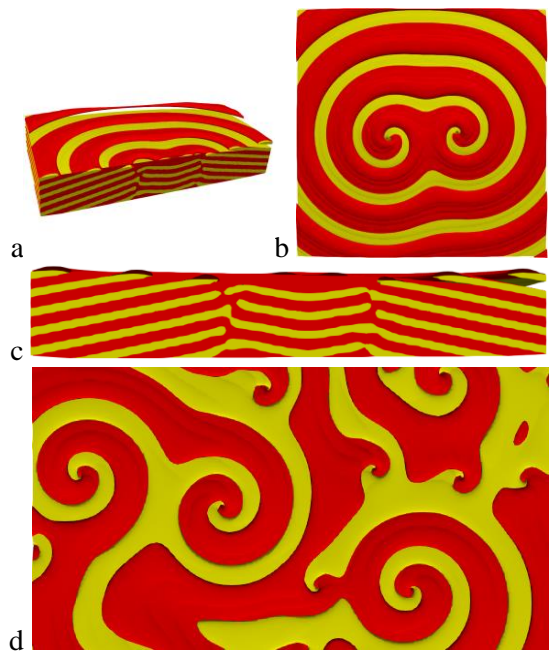

**Fig. 14.** Screw dislocation pair formed in model PF3 at composition **c** = (0.43, 0.55, 0.02). (a) cross section through the axes of the pair. (b) Top view. (c) Front view of the central part of the section shown in panel (a). Multiple screw dislocation pairs formed at **c** = (0.44, 0.54, 0.02) Red and yellow colors indicate solid solution phases rich in the mineral and organic components.

*5.3. Modelling of coral skeletons in model PF1.*

Next, using model PF1, we try to find a qualitative answer to the question, why the skeleton of some corals species contain small randomly oriented crystallites, "sprinkles" (see Fig. 4e), that occur at the perimeter and along grain boundaries, and even form bands, whereas other species do not display this behavior, an observation discussed in some detail in Ref [28]. In our previous work, we demonstrated that conditions of mineralization can influence the amount of sprinkles, however, we have addressed only tangentially the formation of sprinkle bands.

In this section, we address the latter phenomenon. We hypothesize that coral polyp sits on the corallite (top of the skeleton), and creates a supersaturated extracellular calcifying fluid that is not in direct contact with the seawater. Recent work shows that coral skeletons are deposited biologically and actively via attachment of ACC nanoparticles, with ions filling the interstitial space between nanoparticles [123]. It is thus reasonable to assume that two diffusion processes play a role in continuous formation of this thin (< 1 μm) amorphous surface layer: Brownian motion of solid ACC nanoparticles and ion diffusion in the aqueous solution filling the space between polyp and the solidification front. The amorphous surface layer crystallizes slowly into aragonite plumose spherulites. It may be then expected that the skeleton grows into the extracellular calcifying fluid by a mechanism at least partly controlled by these two types of diffusion. Since in the case of coral skeletons we do not need to produce nacre-like alternating organic-inorganic layers via transition from diffusionless to diffusive growth, we can use conditions, where $dv/D_{c,M} \ll 1$, i.e., the system is far from the diffusionless growth regime. As a result, Mullins-Sekerka-type diffusional instability occurs [124], and fingers evolve at the surface of the skeleton, yielding surface roughness on much larger length scale than the thickness of the surface ACC layer (see Fig. 15a-15c). The properties and conditions used in the present simulations for coral skeletons are the same as we used in Ref. [28].

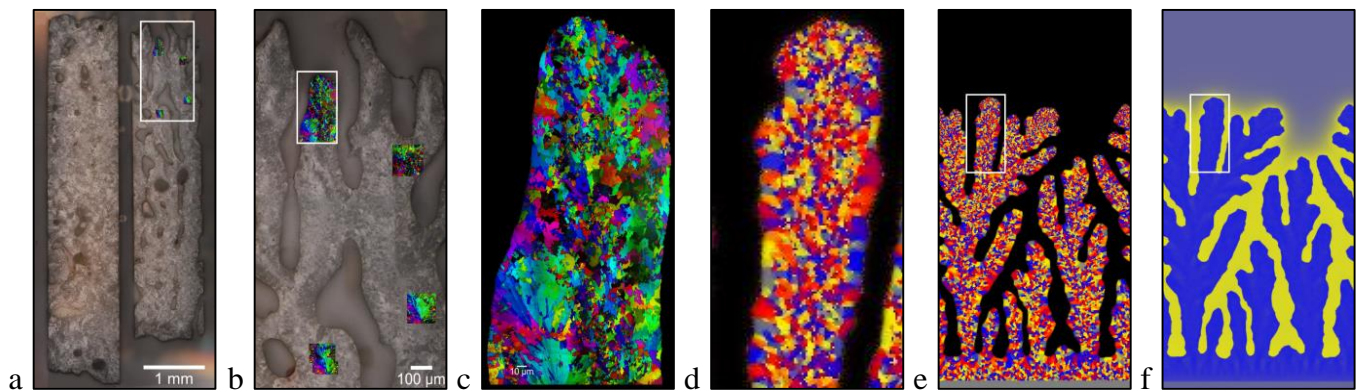

**Fig. 15.** Comparison of the cross sectional microstructure of top part (corallite) of the skeleton of coral species *Balanophyllia europea* (a)–(c) to a simulation performed using by Model PF1. The 2D simulation was performed on a 1000×2000 grid. Orientation maps obtained by PIC mapping are shown in (a)–(c). For comparison, the computed orientation maps are displayed in (d) and (e), whereas the computed composition map is presented in panel (f). In panels (b)–(e) different colors correspond to different crystallographic orientation. The white rectangles in (a) and (b) denote areas magnified in (b) and (c), respectively, whereas the white rectangles in panels (e) and (f) indicate the area shown magnified in panel (d).

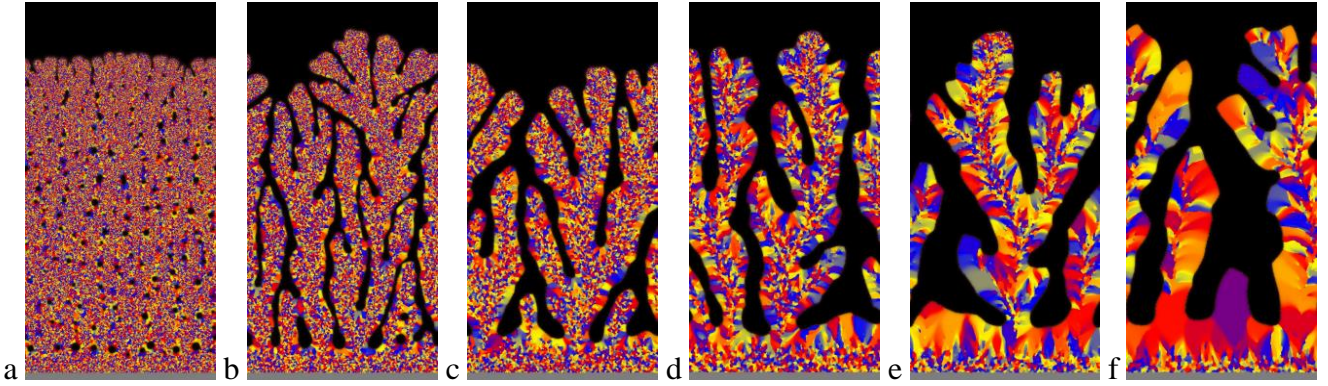

**Fig. 16.** Morphology and microstructure evolution as a function of temperature as predicted by model PF1. The reference computation is shown in panel (d). The relative temperatures from left to right are $\Delta T = -3, -2, -1, 0, 1,$ and $2$ K. The simulations were performed on 1000×2000 grids. Note the decreasing amount of sprinkles with decreasing undercooling (from left to right).

The polycrystalline structure emerging due to the diffusional instability in 2D is shown in Figs. 15d–15f. Note the rough surface and the liquid channels, with small crystallites at the tips and along the spine of the branches, with larger crystallites at the lateral surfaces. This distribution of size is the result of the fact that the growth velocity at the tips is larger due to the larger supersaturation (the tip meets fresh, non-depleted liquid see Fig. 15f), whereas in the inter-arm channels the fluid is depleted, so growth is slow yielding larger crystallites. This phenomenon is the result of growth front nucleation (GFN) that leads to more frequent GFN events with increasing growth velocity as discussed in [68–71, 125]. Whether the combination of diffusional instability with GFN or some other biology directed mechanisms is responsible for the appearance of the rough surface of the skeleton is unclear at present. It is thus desirable to investigate further consequences of the hypothesized control of the size distribution by diffusional instabilities.

Along these lines, we make predictions on the basis of model PF1, which predictions can be used to design experiments that can be used to check the validity of the predictions and via this clarify, whether besides merely reproducing / mimicking the microstructural features, the predictions of the model are indeed in agreement in a broader sense with the experiments.

Accordingly, we investigate factors that influence the amount of sprinkles forming in the model and may offer an explanation why it varies in different species. Previous work indicated that the orientational mobility (related to the rotationnal diffusion coefficient), the thermodynamic driving force (influenced by supersaturation or temperature) may influence the intensity of GFN (formation of new grains at the growth front)

[68, 71, 126]. In the framework of this study, we varied individually these parameters, and in all cases obtained a transition from microstructures dominated by sprinkles to microstructures with few or no sprinkles. Of these parameters only the temperature can be controlled with relative ease. Variation of the supersaturation of the fluid below the coral polyp, or controlling the rotational diffusion coefficient of the ions in the fluid are probably beyond the reach of the experimenter. Therefore, we present only the microstructural / morphological changes predicted as a function of temperature (see Fig. 16). Apparently, according to our model, the coral skeletons grown at low temperatures should have larger amount of sprinkles than those grown at higher temperatures. This finding raises the question whether the amount of sprinkles is indeed a characteristic feature of the individual coral species (i.e., determined biologically) or some differences in the circumstances of mineralization (temperature and/or supersaturation) are responsible for the deviations. In any event, these simulations may offer a natural explanation for the differences seen in the amount of sprinkles in the skeleton of different coral species.

## 6. Summary and concluding remarks

We have demonstrated that coarse-grained phase-field modeling offers a methodology to address specific mesoscale aspects of non-classical crystallization phenomena taking place during biomineralization. In particular, we investigated how far phase-field modeling can contribute to the understanding of microstructure evolution during the formation of mollusk shells and coral skeletons for various species. We applied three different phase-field models: PF1, PF2, and PF3.

Our present findings can be summarized as follows:

(1) *Ultrastructure specific to the shells of mollusks Unio pictorum, Nautilus pompilius, and Haliotis asinina*: Binary phase-field models PF1 and PF2 recover the common sequence of granular → prismatic → nacre ultrastructures on a reasonable time scale, if CC crystallization takes place via an amorphous precursor. In contrast, CC crystallization via ion-by-ion de-position from aqueous solution appears to be orders of magnitude too fast when compared to experiments.

(2) *Nacre formation in mollusk shells:* Models PF1 and PF2 describe reasonably well the formation of not

only the granular and prismatic domains, but the appearance of sheet nacre as well, in which case the models indicate alternating precipitation of the organic and mineral components. The models seem to reproduce even such details as mineral bridges and aligned holes. Yet, for obvious reasons, they cannot predict the formation mechanism of columnar nacre, in which the formation of organic membranes precedes CC precipitation. However, representing the preexisting organic membranes via appropriate boundary conditions, a reasonable description can be obtained even in this case.

(3) *Screw dislocations in mollusk shells:* Ternary phase field model PF3 predicts the formation of screw dislocations pairs in 3D, a phenomenon analogous to the experimental findings. Inclusion of elasticity into the model is needed to capture the proper dynamic behavior during growth.

(4) *Sprinkle formation in coral skeletons:* Model PF1 was used to explore possible mechanism for the formation of nanoscale crystallites "sprinkles" whose presence was recently in the skeletons of certain coral species. Assuming a diffusion controlled mechanism in confined space, we observe the formation of sprinkle bands at the spine of the arms of the corallite as a trace of fast solidification at the arm tips, whereas larger crystallites form at the sides of the arms. The simulations show that varying the orientation mobility (proportional to the rotational diffusion coefficient of the molecules / ions) or the driving force of crystallization (via changing either the supersaturation or the temperature) one can control the amount of sprinkles between essentially no sprinkle and dominantly sprinkled microstructures.

Unquestionably, the applied models should be viewed as only minimum models of the processes taking place during biomineralization. Yet, we believe that phase-field modeling complemented with biochemical / biological information can contribute to a better qualitative or perhaps even quantitative understanding of morphogenesis in simple cases. Introduction of more complex models can certainly improve the mathematical representation of the associated phenomena. Ultimate limitations of such approaches stem from the fact that living organisms cannot be modeled within this framework: they can only be represented by boundary conditions of different complexity. Despite these, in specific cases we recovered structures closely resembling their biogenic counterpart. The resemblance of the simulated and natural biominerals suggests that, underneath the immense biological complexity observed in living organisms, the underlying design principles for biological structures may be so simple that they can be understood with simple math, and simulated by phase-field theory.

Finally, we note that our simulations outline conditions, under which standard materials science processes can be used to create inorganic substances that mimic the microstructures observed to form in living organisms, a knowledge that may open up ways for creating new biomimetic / bioinspired composite materials.

**Acknowledgments**

This work was supported by the National Agency for Research, Development, and Innovation (NKFIH), Hungary under contract No. KKP-126749. Eutectic codes used in this work were partly developed in the framework of the NKFIH contract No. NN-125832. Research infrastructure in Hungary was provided by the Hungarian Academy of Sciences. PG received 80% support from the U.S. Department of Energy, Office of Science, Office of Basic Energy Sciences, Chemical Sciences, Geosciences, and Biosciences Division, under Award DE-FG02-07ER15899, and 20% support from NSF grant DMR-1603192.